\documentclass[pre,twocolumn,amsmath,amssymb,floatfix,superscriptaddress]{revtex4-2}

\usepackage{graphicx}
\usepackage{dcolumn}
\usepackage{bm}
\usepackage{booktabs}

\usepackage{lineno}

\usepackage{lipsum}

\usepackage[FIGTOPCAP,tight,raggedright,nooneline]{subfigure}
\usepackage{float}
\graphicspath{{images/}{plots/}}

\usepackage{xcolor}

\definecolor{applegreen}{rgb}{0.55, 0.71, 0.0}

\usepackage[normalem]{ulem} 
\usepackage{hyperref}
\usepackage{url}

\begin{document}

\title{Who talks about what? Comparing the  information treatment in  traditional media with  online discussions}


 \author{Hendrik Schawe}
    \email{Hendrik.Schawe@gmail.com}
    \affiliation{Laboratoire de Physique Th\'{e}orique et Mod\'{e}lisation, UMR-8089 CNRS, CY Cergy Paris Universit\'{e}, France}

  \author{Mariano Gastón Beiró}
    \email{mbeiro@fi.uba.ar}
    \affiliation{Universidad de Buenos Aires. Facultad de Ingeniería, Paseo Colón 850, C1063ACV Buenos Aires, Argentina}
    \affiliation{CONICET, Universidad de Buenos Aires, INTECIN, Argentina}
  
    \author{J.Ignacio Alvarez-Hamelin}
    \email{ihameli@cnet.fi.uba.ar}
    \affiliation{Universidad de Buenos Aires. Facultad de Ingeniería, Paseo Colón 850, C1063ACV Buenos Aires, Argentina}
    \affiliation{CONICET, Universidad de Buenos Aires, INTECIN, Argentina}

    \author{Dimitris Kotzinos}
    \email{dimitrios.kotzinos@cyu.fr}
    \affiliation{ETIS UMR 8051 CY Cergy Paris Université, ENSEA, CNRS, France}

    \author{Laura Hern\'{a}ndez}
    \email{laura.hernandez@cyu.fr}
    \affiliation{Laboratoire de Physique Th\'{e}orique et Mod\'{e}lisation, UMR-8089 CNRS, CY Cergy Paris Universit\'{e}, France}

    \date{\today}

    \begin{abstract}
    We study the dynamics of interactions between a traditional medium, the New York Times journal, and its followers in Twitter, using a massive dataset. It consists of the  metadata of the articles published by the journal during the first year of the COVID-19 pandemic, and the posts  published in Twitter by a large set of followers of the @nytimes account along with those published by a set of  followers of several other media of different kind. 
     The dynamics of discussions held in Twitter by exclusive followers of a medium show a strong dependence on the medium they follow: the followers of @FoxNews show the highest similarity to each other and a strong differentiation of interests with the general group. Our results also reveal  the difference in the attention payed to U.S. presidential elections by the journal and by its followers,  and  show that the topic related to the ``Black Lives Matter'' movement  started in Twitter, and was addressed later by the journal.

    \end{abstract}

    \maketitle

    \section{Introduction}

The debate about the influence of mass media on social opinion 
 has shown  peaks of interest each time that a technological breakthrough modified the media ecosystem, mainly by increasing the amount of people that can be reached by broadcasters~\cite{10.2307/1019970}. The first important one, the invention of the printing press by Gutenberg,  has indeed  played an  important role in the rapid expansion of Calvinism in Europe~\cite{248510}, although its general influence on the formation of social opinion  was mitigated by the fact that, most of the population  was  illiterate.  
 Later, about the beginning of the 20th  century, when the wireless  radio transmissions appeared and rapidly became a popular entertaining medium, discussions about the foreseeable consequences of the popularization of this new medium were carried in the written press, which by that time had become a traditional one. A review in the New York Times from May 7th 1899 entitled  ``Future of Wireless Telegraphy'' warned:
\textit{``All the nations of the earth would be put upon terms of intimacy and men would be stunned by the tremendous volume of news and information that would ceaselessly pour in upon them''}~\cite{url_1899}. 
Needless to say that the same kind of debates took place at the arrival of TV broadcasting ~\cite{10.2307/2991719}.

The rapid growth of digital media certainly triggered again the same kind of discussions but this time, with a major difference: the massive data accumulated on social media platforms allows us to  perform measurements about the opinion evolution of large amounts of people. 
A countless number of articles,  have addressed different aspects of opinion dynamics based on social networks. A few recent ones, are  the study   of opinion evolution on different selected topics~\cite{gaumont2018reconstruction,boutet2013s},  the characterisation of structural properties of the   interaction networks that result from the different functionalities  offered by the platforms  (like mentions, retweets, follower-friend in Twitter)~\cite{himelboim2013tweeting,barbera2015birds}, with a recent  particular interest on the formation of \textit{information bubbles} and \textit{echo chambers} -strongly connected clusters of people that communicate only weakly with others-~\cite{ScienceOpenVid:2cb18dc9-e556-4c25-bcae-c0751724fde6,cinelli2021echo,choi2020rumor}. Special attention has been given to the diffusion of rumours and fake news in relation with the COVID-19 pandemic~\cite{Lazer_fake}, to the extent that the term \textit{infodemics} was coined to highlight the parallelism with the  diffusion of the virus~\cite{infodemics,menczer2021covid,shahi2021exploratory}.

Nowadays, it seems clear that if  media exert an influence on social opinion it is  mainly by setting the terms of debate or, in the words of B. Cohen~\cite{cohen2015press}, \textit{  the press may not be successful much of the time in telling people what to think, but it is stunningly successful in telling its readers what to think about.} This notion is known as the \textit{Agenda Setting Problem}~\cite{10.1086/267990}. 

In this work we investigate the agenda setting problem,  by studying the dynamics of the different topics treated by a traditional medium, \textit{The New York Times} (NYT) journal, in relation  with  the dynamics  of the public discussion among its followers on  Twitter. We center our study in the first year of the COVID-19 pandemics, which by its very nature, can be expected to become an important driver of public attention.
Several 
works studied the evolution of the opinion in Twitter (and other platforms)  during this period, mainly focusing on  discussions directly related to health issues or public policies related with them  ~\cite{SACCO2021114215,cinelli2020covid, collection_covid}.   

Here, on the contrary,  we 
aim at understanding how the different topics that interested the society during this period were addressed both by the media and by the public that is  in direct relation with them, without assuming a priory  the existence of any influence on either direction.
 
 While some recent  studies have compared how  traditional media and social networks  treat \textit{a particular topic} of discussion~\cite{vargo2015event,morris2018twitter,bridgman2021infodemic,su2019agenda,ceron2014twitter},  here we search for global patterns characterizing each of them.  

We have collected a large amount of tweets corresponding to a randomized sample of the over 46M followers of the New York Times (NYT) official Twitter account (@nytimes), during the first year of the pandemic, along with the metadata of the articles published by the journal during that period. This sampling guarantees that we are reaching the topics discussed by users that have expressed an interest in that journal by following its Twitter account. 
In order to compare with the behaviour of the followers of different media, we have also collected a sample of the tweets published by the followers of other important media of different kinds: written press, radio, television, press agencies.

With this data, we  build a semantic network representative of the discussion taking place in social media, based in the co-occurrence of \textit{hashtags} -tagging words starting with the symbol ``\#''-, chosen by Twitter users. By community detection on this semantic network, we identify the topics of interest discussed in the platform. On the other hand, the   keywords chosen by the journalists to tag their articles allow us to identify the topics treated by the journal. 

 By an extensive analysis of these data   we aim at getting an insight into the following questions:
 \begin{itemize}
 \item Who talks about what? Do people that follow a journal talk about the same subjects that are published in the journal? In that case,  is it possible to quantify to what extent?
 \item How the attention that the followers of the journal pay to different topics, compares to the attention that followers of other media pay to those topics?
 \item Can we observe any evidence about the agenda setting problem? If so,  in which sense?
 \end{itemize}

    \section{Results}
        \label{sec:results}

   We collected data for over a year, starting on January 2020,  before the outbreak  of the pandemic, from  followers of the @nytimes in Twitter,   and also from  Twitter users who follow  Twitter accounts of other media,   like @washingtonpost,  @WSJ (The Wall Street Journal), @TIME,  continuous information television channels, like @FoxNews, or @CNN and also press agencies like @AP  (Associated Press) or @Reuters. During the same period, we have also collected  the metadata of the publications of the NYT journal, in particular the articles' headers (see Methods).    

In order to automatically determine the topics of discussion in Twitter,  we build a hashtag network where two hashtags are connected if they appear in the same
tweet (see Methods). This link is weighted by the number of \textit{different} users that used  that pair of hashtags,  which diminishes the potential influence of automated accounts. 

This semantic network relies on a single  assumption:  if two hashtags appear in the same tweet, they are likely to refer to the same subject.  As a given subject may be addressed to by different hashtags,  the topics of discussion in the platform are automatically obtained by   community detection in the semantic network~\cite{arg_elections, cardoso2019topical}, and we consider that each community constitutes a \textit{topic} of discussion in the platform (see Methods). 

The topics  treated in the NYT articles, are labelled by  the keywords given by the journalists to characterize each article.

    \subsection{Topic  dynamics}
    \label{subsec:topics_dyn}
    
    The entropy of the vocabulary (hashtags for Twitter and keywords for the NYT journal) allows for a global comparison  of the dynamics of the discussion in both media (see  Methods). Low values  indicate that the discussion is concentrated around few hashtags or that the information can be tagged by few keywords, while high values reveal that many different  hashtags or keywords enter in the discussion.  
  
\begin{figure*}[t]
            \centering
            \includegraphics[scale=1]{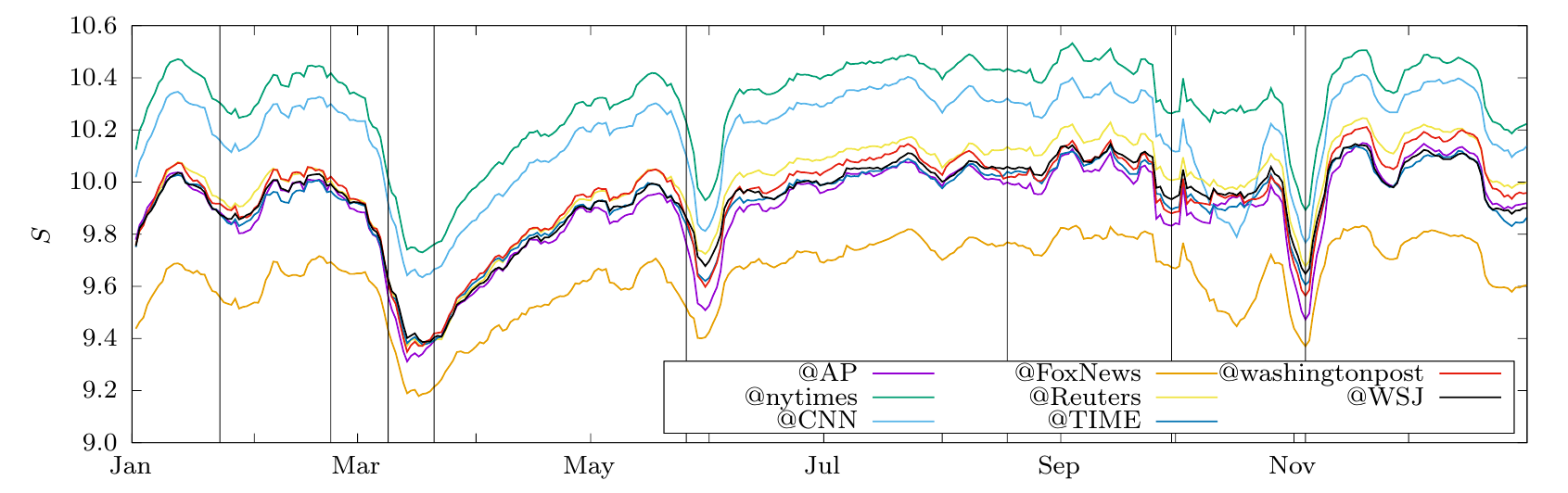}
            \includegraphics[scale=1]{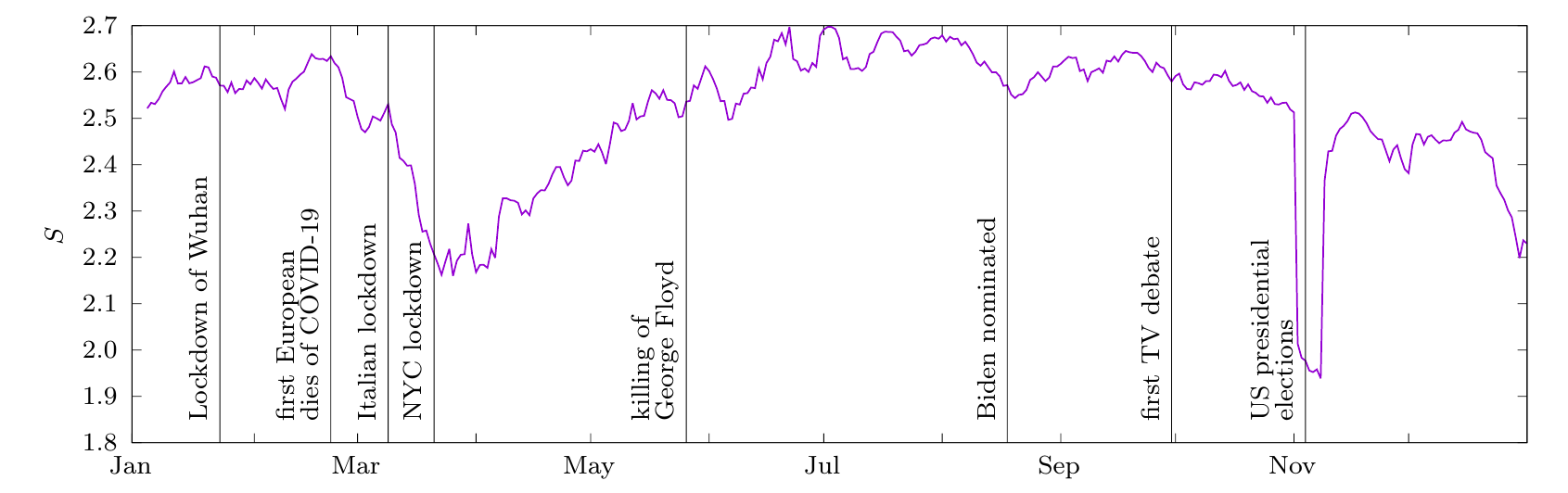}
            \caption{\label{fig:events}
                Entropy of the vocabulary  as a function of time. Top panel: Time variation of the hashtags' distribution entropy corresponding to the frequencies of   usages  of each hashtag by  the Twitter followers of the accounts of the media listed in Tab.~\ref{table_media}. Bottom panel: Time variation of the keywords' distribution entropy corresponding to the keyword usages in the articles of the New York Times. The vertical lines indicate as a reference, the time location of important events during the studied period. In both cases the computation has been done daily with a 7 day rolling averaging (3 days before, 3 days after) to remove the effect of weekdays.
                            }
        \end{figure*}

   Fig.~\ref{fig:events} (top) shows   the entropy of the hashtags used by 
   the followers of different media. The dates of important events are marked as temporal references.

        As the entropy is an extensive variable, it is not surprising to see that the corresponding values are, in general, larger for Twitter than for the NYT, 
         since the number of hashtags is much larger than the number of keywords. 
         
       Also, the entropies corresponding to the  @nytimes and @CNN followers, who are significantly more numerous than those following  the remaining media accounts, are globally larger than the other curves, as more users naturally lead to more hashtag usages.  
         
        The unexpected, opposed situation is  observed  for the  @FoxNews followers, whose entropy is always the lowest  in spite of the fact that this is not the smallest group, revealing that @FoxNews followers use fewer hashtags than expected by their number.
      This is not due to a particular event that could have interested these followers, but it is  constant in time, which indicates a characteristic of those users.  
  
  Fig.~\ref{eq:entropy} (bottom) shows a different dynamics for the evolution of the entropy of keywords of the NYT journal.   Although the publications in  both supports are naturally attached to real life events,  a detailed inspection of 
   the most popular topics in both media confirms also structural differences. For instance,  the discussions about the  `Black Lives Matter movement', notably gives an earlier signal in the entropy of Twitter, while its influence is hardly detectable after the killing of George Floyd in the entropy of the NYT keywords. This observation may be related to the fact that, unlike Twitter,  a journal follows editorial policies  which mostly lead to a balanced reporting about different topics.

As expected, the `Coronavirus topic dominated both online discussions and also the journal articles , capturing the attention during a long period, as shown by  the wide entropy decrease in March-April. As COVID-19 influences most aspects of life, it  appears in many sections of the journal  and    has considerable influence on the entropy.
   
The `Presidential Elections' topic,  visible in both entropy panels, shows a steeper valley for the NYT curve (deeper than the dominating `Coronavirus' topic). 
   This reflects the importance of the covering of elections by the journal, as NYT publishes, among others, one article of the election results for each of about 400 districts of the United States, really focusing on the subject during this period.

        \begin{figure*}[htb]
            \centering
            \includegraphics[scale=1]{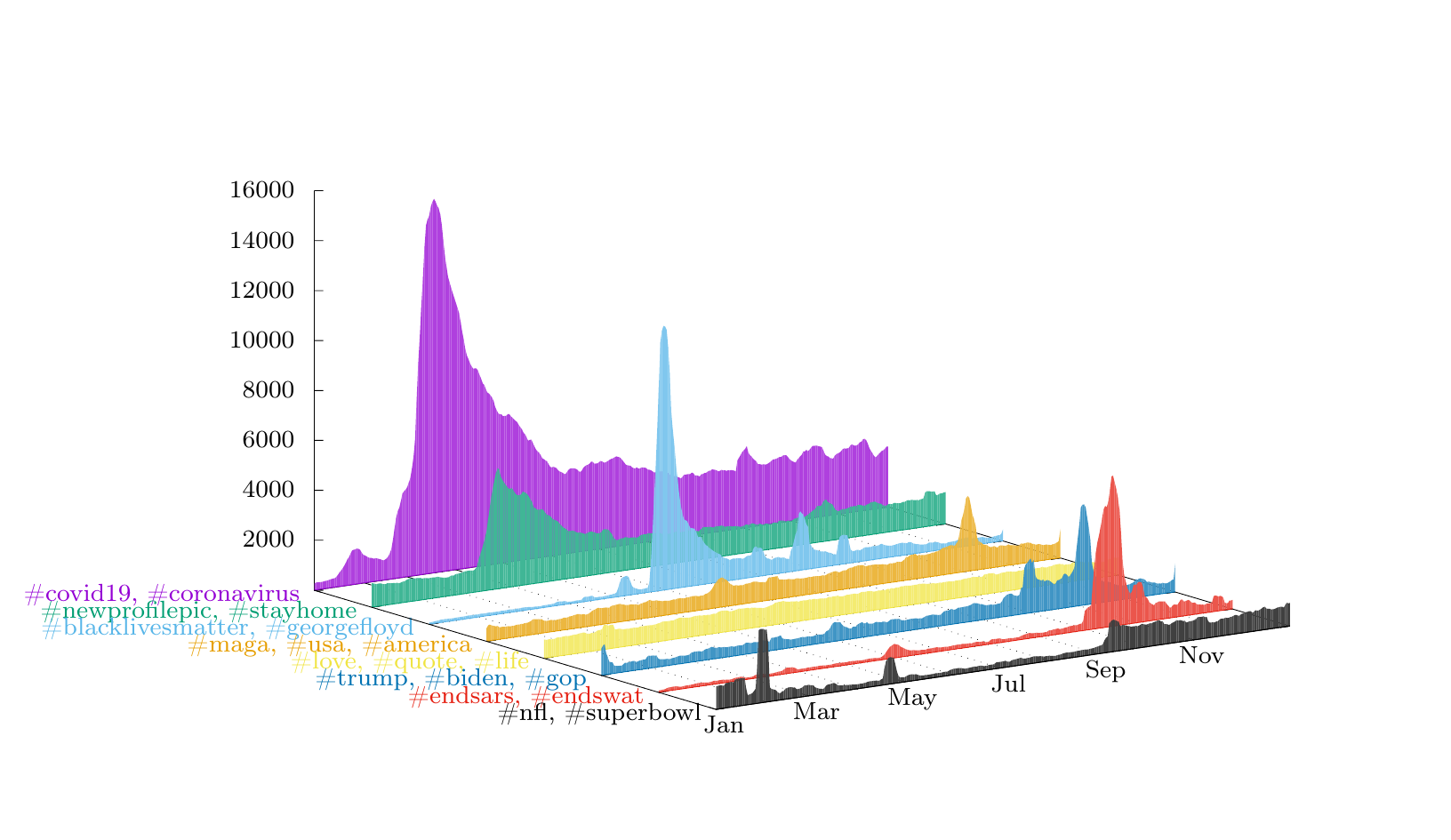}
            \caption{\label{fig:topics_dynamics}
                Dynamics of the eight largest topics of the discussion in Twitter by the followers of @nytimes account.  The topics are identified in the semantic network of co-occurring hashtags by Infomap (one level down, i.e., path length of 2). They are labelled by the most used hashtags belonging to each topic. The vertical axis shows the number of unique users using a hashtag belonging to the community of the corresponding topic on a given day, smoothed with a rolling average over seven days to eliminate the cycles introduced by weekends.
            }
        \end{figure*}
        
        A closer look at the topics' evolution, allows us to confirm that the remarkable decrease observed in the entropy curves around the dates of important social events, are indeed caused by topics related to them.
     
   Fig.~\ref{fig:topics_dynamics}    shows the evolution of the eight most popular topics which are labelled by the most used hashtags in the corresponding community.
   
   We can also detect  the effect that the pandemic had on the public discussion  about subjects that seem a priori completely unrelated to it.  For example, the topic labelled by the hashtag  
 \#newprofilepic, includes other  hashtags related to locations and also the hashtag  `\#flashbackfriday' which is used to tag pictures. 
We find that this topic becomes connected to the coronavirus pandemic via the \#stayhome hashtag, presumably because of the changing nature of pictures posted under these hashtags due to lock-down period. 
 
 The fact that  our method  lets the topics to  emerge, instead of following   a   set  of hashtags or keywords   chosen a priori, reveals interesting facts.  We find a  topic whose popularity may appear as  surprising in US society,  labelled by the `\#endsars'   hashtag. In fact, this topic refers to the demonstrations against police violence sparked by videos showing brutality of the Nigerian police organization SARS (Special Anti-Robbery Squad, not to be confused with the SARS-Coronavirus). After detecting the users who talked about this subject we found that most of their accounts were tagged abroad (see Supplementary Material).

        \begin{figure*}[t]
            \centering
            \includegraphics[scale=1]{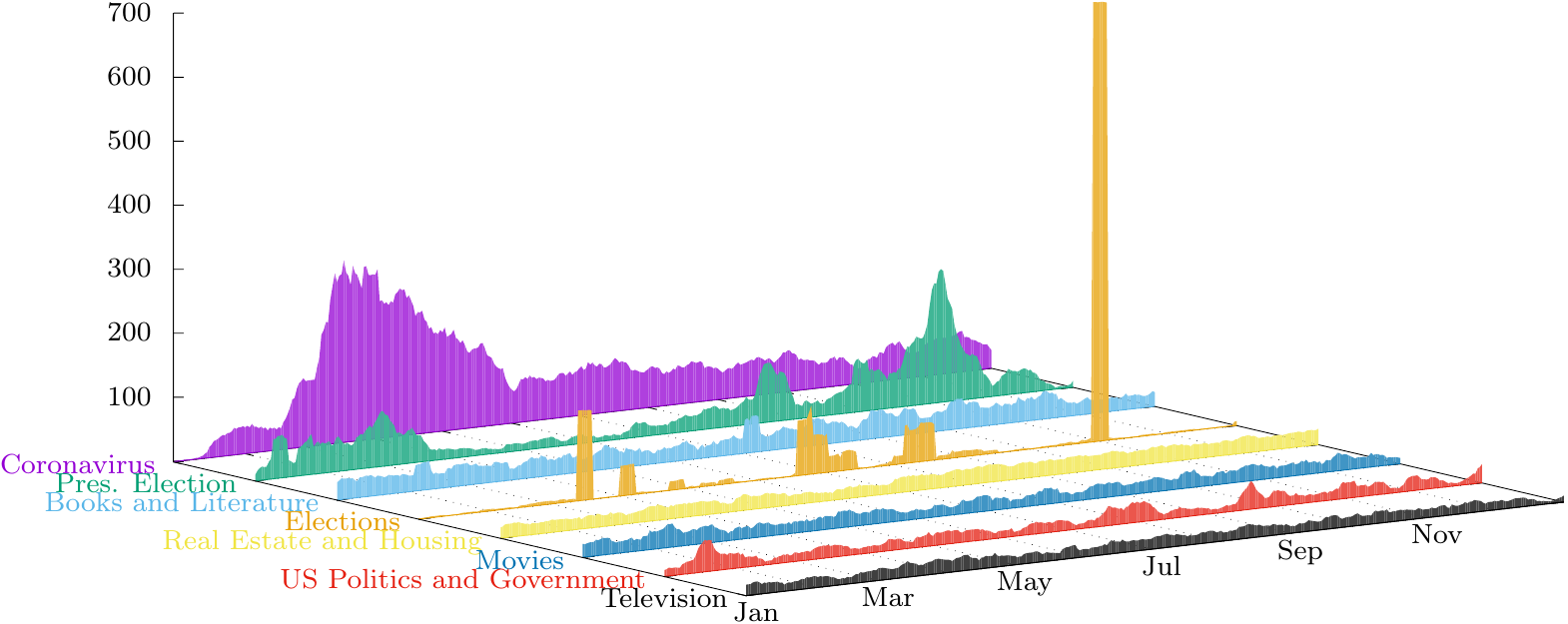}
            \caption{\label{fig:kw_dynamics}
                Dynamics for 8 most used primary keywords tagging the NYT articles (shortened names used in the labels). The vertical axis shows the number of primary keyword usages on each day, smoothed with a rolling average over 7 days to eliminate structure introduced by weekends (see Supplementary Material).
            }
        \end{figure*}
        
 The  dynamics of the topics treated by the  NYT journal,  is shown in Fig.~\ref{fig:kw_dynamics}. As for Twitter, we also observe topics reporting  events, like `Presidential Election' or  `Coronavirus' and those which correspond to regular reporting of different aspects of social life like `Books and Literature'. 
   As signaled by the entropy curves, the `Black lives matters' topic which dominates in Twitter (cf.  Fig.~\ref{fig:topics_dynamics}) is not even among the leader keywords depicted here. On the contrary  the `Presidential Election' and `Elections' keywords, which also includes articles about  the results of the presidential election for all districts, shows that this subject dominates the journal attention even in the background of the pandemics, while it is much less important for its followers which refer to it by the topics labelled by \#maga and \#trump.

   \subsection{Rank diversity}
    \label{subsec:rank_div}

        By definition (see Methods), the rank diversity   has  low value  when few hashtags or keywords have occupied  the observed rank during the corresponding period. For the first ranks, this low  value reflects that the leading subjects were represented by few hashtags or keywords.

\begin{figure}[t]
            \centering
            \includegraphics[scale=1]{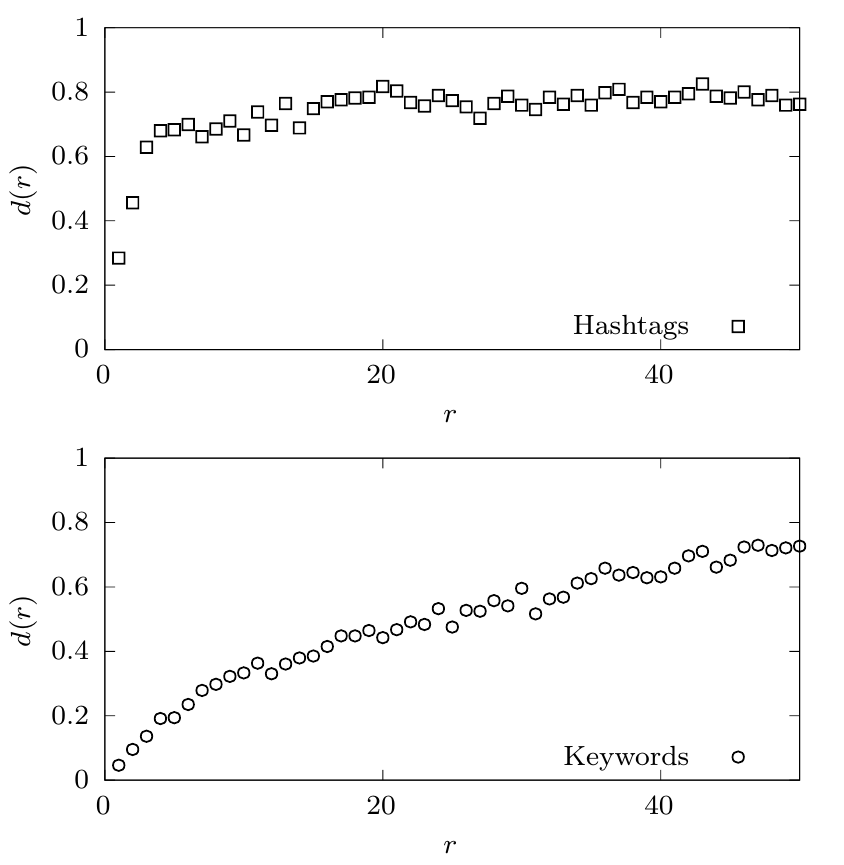}
            \caption{\label{fig:rank}
                Daily rank diversity $d(r)$ for the 50 most used hashtags (top panel) and keywords (bottom panel).
            }
        \end{figure}

       This measure has been introduced to study the evolution of languages, which takes place over long periods,  \cite{cocho2015rank} by analysing the evolution of rank of single words or n-grams \cite{Morales2018rank}. 
     The  plots of Fig.~\ref{fig:rank},which concern a  much shorter time-scale,   differ from the sigmoid curves characteristic of language evolution.  In Twitter and in this particular period, one could have expected to have  first ranks completely dominated the few variations of   COVID-19 hashtags.   However,  this is not exactly the case. Only rank one and two have a diversity below $d(r) < 0.5$, which indicates that there are many  different hashtags occupying those ranks once over the year ($d(r) =0.5$ correspond to  $180$ different hashtags.See Methods).
       This doesn't necessarily mean that the users were ignoring the  pandemic in their discussions but, as  it  affected very different aspects of society,   it can be addressed to by many different hashtags. In fact, we have found that the COVID-19 topic is composed of about $700$ hashtags (see Supplementary Material), several of which are very popular and contribute to the relative variability of the first ranks.

       The rank diversity   clearly captures the structural difference between these two media,   showing a completely different shape for the keywords of the NYT journal.  Since the keywords are curated, the bottom panel of  Fig.~\ref{fig:rank} reveals that dominant topics of the journal are addressed by very few keywords. This shows that the
       NYT has a narrower focus on a selected group of topics than their followers on Twitter.

    \subsection{Reaction of the followers of NYT to  its publications }
    \label{sec:delay}
     Here we investigate the patterns that characterize the reactions of the   followers of  the @nytimes account  to the articles and tweets published by the journal.    
     
     We observe two different kind of reactions: a \emph{direct} one takes place directly on  the Twitter platform,    when the followers  retweet, quote or reply to the tweets published by the journal's account. Another \emph{indirect} reaction takes place  by the means of the `share on Twitter' button of the website \url{nytimes.com} where the followers of the journal can tweet a link to the article of their choice.
   
\begin{figure}[t]
            \centering
            \includegraphics[scale=1]{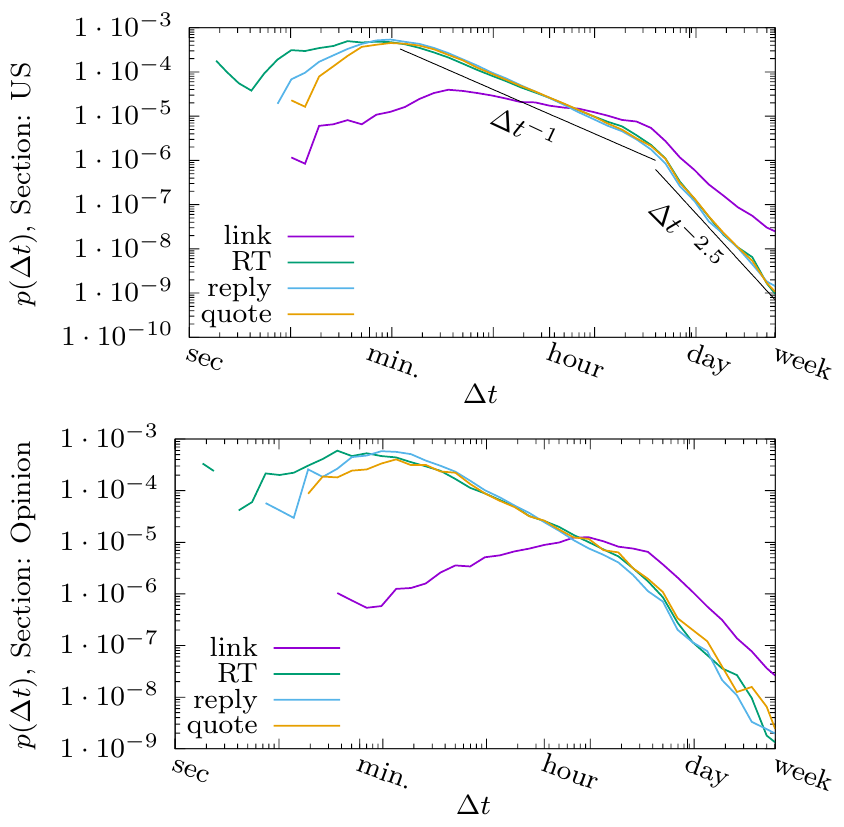}
            \caption{\label{fig:delay_op_us}
                Distribution of the delay times, $\Delta t$: Elapsed time between the direct reaction of the users (retweet, reply or quote) and the issue of  a tweet from the @nytimes account  or between  the indirect reaction, retweet of an  the article via the ``share'' button in the website, and the appearing of the article in the journal (violet lines, labeled \emph{link} in the figure.) 
               We show  here the distributions  corresponding to two sections of the journal in which the corresponding articles appeared. Top:  ``U.S.'' section Bottom: ``Opinion'' section (for other sections, see Supplementary Material).
            }
        \end{figure}

   These two kind of interactions between the journal and its followers are characterized by  different regular  patterns.   Figure~\ref{fig:delay_op_us} shows the distribution of  \textit{reaction times}, $\Delta t$, the delay between either the tweet of the @nytimes account and the direct reaction of the user, or the delay between the publication of an article online and the tweet published by the follower using the website button.

  The first observation is the  very broad range spanned by the reaction times, going from seconds to a week. There is a 
   striking difference in the shape of the curves corresponding to direct reactions, where the distribution of reaction delays   seem to fit a power-law with a breaking of the slope around $10$ hours, and that of indirect reactions  which are much slower and start in general by a very broad shallow  peak followed by a power law decrease again around $10$ hours.
   
   The numerous retweets happening within the first second of the original tweet, suggest the presence of   automated users. The qualitative behaviour of the reaction times is the same for the three direct reactions curves.
Fitting a power law after the maxima of the distributions, we find   a breaking of the slope from $\Delta t^{-1}$  to a fastest decrease, $\Delta t^{-2.5}$.

  The extremely similar behavior of all direct interactions suggests that this process may  strongly be influenced by the way in which the platform presents the tweets of followed users, where older tweets are pushed out quickly from a user's timeline by newer tweets. In this case, the first retweets of an article may trigger more retweets from the users that might have lost the original tweet from @nytimes in their timeline, in a manner of a self exciting process~\cite{hawkes1971spectra,rizoiu2017tutorial}.

      The longer reaction times observed for indirect reactions are  expected, assuming that followers which are on the NYT website are more likely to read the article before sharing it, such that the most probable reaction time is shifted to multiple minutes or hours. Also here, we observe a strong decrease at the $\approx 10$ hour mark. An important difference with the  direct reactions is that   the distribution of response times for link sharing does not look universal, showing a different shape for different sections (see Supplementary Material)

        Despite the extremely similar shape of the delay distributions of the direct Twitter interactions, the median  delay time fluctuates by more than a factor of two for different sections, as shown in  Table \ref{tab:delays}.  
        Links to articles about books and art  are posted for longer time (median delay of over one day), than national (U.S.) or international (World) news (median delay of about half a day), which seems expected considering that book reviews should remain of interest for  longer times than the typical everyday news item. However,  the behaviour of the   direct  reactions shows the  opposite tendency: Books and Art are the sections with the shortest median delays before retweets, while national and international news are amongst the slowest sections regarding retweets.

        \begin{table}[t]
            \begin{ruledtabular}
                \begin{tabular}{rllll}
                    Section
                        & \multicolumn{1}{c}{${\Delta t_\mathrm{RT}}$} 
                        & \multicolumn{1}{c}{${\Delta t_\mathrm{reply}}$} 
                        & \multicolumn{1}{c}{${\Delta t_\mathrm{quote}}$} 
                        & \multicolumn{1}{c}{${\Delta t_\mathrm{link}}$}\\[0.05cm]
                    \hline \noalign{\vskip 0.1cm}
                    U.S.         &  $\phantom{1}81.9(5)$ & $52.3(6)$ &  $\phantom{1}71.0(9)$  &  $\phantom{1}643(3)$  \\
                    World        &  $\phantom{1}90.5(9)$ & $45.9(7)$ &  $\phantom{1}73.4(17)$ &  $\phantom{1}827(6)$  \\
                    Opinion      &  $\phantom{1}68(3)$   & $38(2)$   &  $\phantom{1}83(7)$    &  $\phantom{1}952(4)$  \\
                    Arts         &  $\phantom{1}53(2)$   & $37(2)$   &  $\phantom{1}46(2)$    &            $1526(18)$ \\
                    Business Day &  $\phantom{1}75(1)$   & $49(1)$   &  $\phantom{1}89(4)$    &  $\phantom{1}814(7)$  \\
                    Sports       &  $\phantom{1}42(2)$   & $27(2)$   &  $\phantom{1}44(4)$    &  $\phantom{1}713(15)$ \\
                    New York     &  $\phantom{1}85(1)$   & $45(1)$   &  $\phantom{1}73(2)$    &  $\phantom{1}597(6)$  \\
                    Books        &  $\phantom{1}45(2)$   & $36(3)$   &  $\phantom{1}40(5)$    &            $1646(28)$ \\
                    Style        &            $103(5)$   & $50(3)$   &            $107(7)$    &            $1065(19)$ \\
                    Movies       &  $\phantom{1}53(3)$   & $34(2)$   &  $\phantom{1}52(4)$    &            $1259(35)$ \\
                    Real Estate  &  $\phantom{1}29(4)$   & $40(7)$   &  $\phantom{1}62(20)$   &            $1735(38)$ \\
                    \hline \noalign{\vskip 0.1cm}
                    all articles &  $\phantom{1}76.7(3)$ & $47.0(3)$ &  $\phantom{1}67.8(5)$  &  $\phantom{1}858(2)$  \\
                \end{tabular}
            \end{ruledtabular}
            \caption{\label{tab:delays}
                Median of the delay $\Delta t$ in minutes for different sections and different types of interactions shown for
                the eleven largest sections sorted by decreasing number of articles assigned to the sections.
                Despite the shape of the distribution being very similar (see Supplementary Material), the median 
                delay fluctuates by more than a factor of two depending on the section. The number in parentheses specifies                 the standard error in units of the least significant digit, obtained via bootstrap resampling \cite{young2015everything}.
                            }
        \end{table}

        We only consider those tweets that reply directly to the original tweet of @nytimes and not replies to other replies, therefore the shorter median delay observed for replies is not caused by fast back and forth discussions.

    \subsection{Characterization of the users}
\label{similarities}
       
             The dynamical study of the discussion taking place in Twitter during the considered period shows that some groups of  users synchronize  in phase or in  anti-phase at some particular moments, revealing that  most of them are discussing about the same subject or talking about completely different ones, respectively.

Here we have determined the communities (topics) in  a  semantic network that also includes  the tweets of the followers of all considered media to  ensure that we do not miss topics which might not be important for followers of @nytimes, but relevant for followers of other media.

       Each user is characterized by a \emph{dynamical topic vector} whose dimension  is equal to  the total number of detected topics. Each component of this vector indicates   whether the corresponding user has tweeted more or less than the average population about  the corresponding topic, as a function of  time (see Methods). 
       
\begin{figure}[t]
            \centering
            \includegraphics[width=\linewidth]{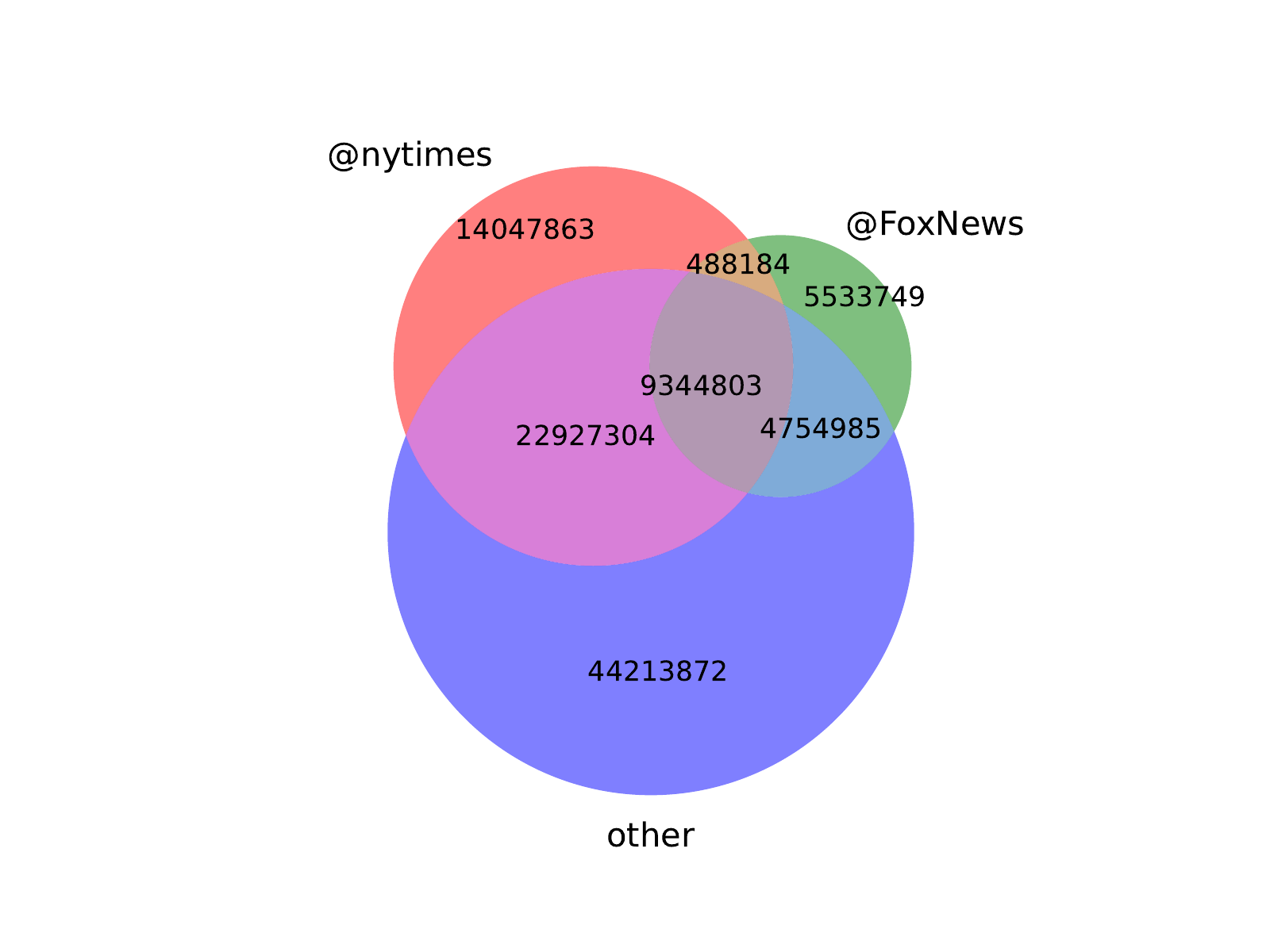}
            \caption{\label{fig:venn}
                Venn diagram of ``followship'' relations  showing the intersection among the followers  of  @nytimes (pink) and @FoxNews (green) and the other media (blue). 
            }
        \end{figure}

       Users are divided into groups according to the media they are following, among the 10 most followed media in US, listed  in Table~\ref{table_media} (see Methods). Figure~\ref{fig:venn} shows that   many users, follow more than a single  medium.   Users following a single medium are called \emph{exclusive followers}. 
       
        Most of the  media  considered here hold a  neutral or liberal position on the political spectrum with a  similar  entropy of their vocabulary,   as shown in Fig.~\ref{eq:entropy}; the exception being @FoxNews, which is considered politically conservative and whose entropy is the lowest as discussed in Section~\ref{subsec:topics_dyn}.

\begin{figure}

            \centering
            \includegraphics[scale=1]{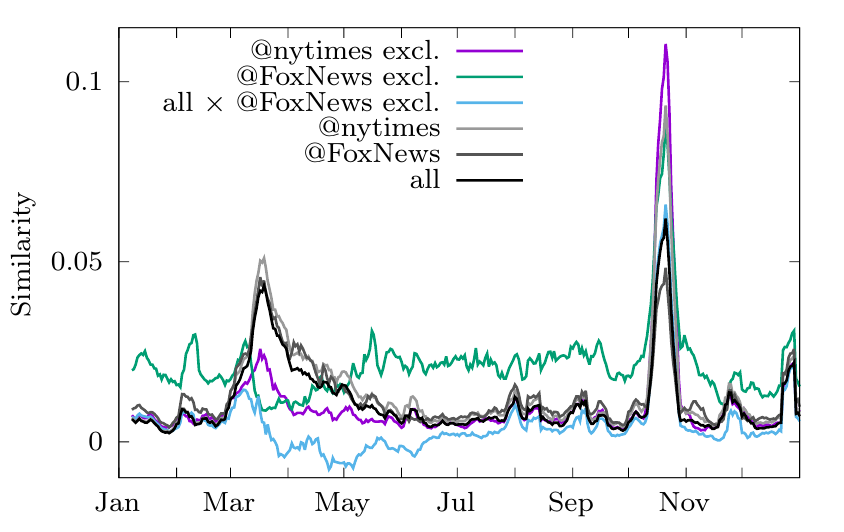}
            \includegraphics[scale=1]{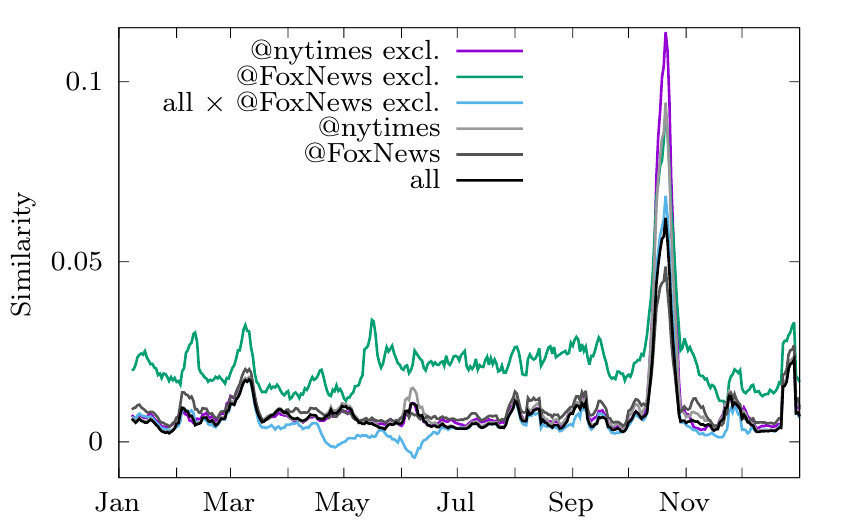}

            \caption{\label{fig:similarity_nyt_fox}
            Dynamics of self and cross-similarities  corresponding to  sub-populations that follow  different  media accounts in Twitter. For clarity we concentrate on the curves involving  the followers of @nytimes and  @FoxNews, along with a randomized sample that includes followers of all media, labelled ``all'' (more curves in the Supplementary Material). The labels `@nytimes excl.' and `@FoxNews excl.' refer to the sub-populations whose members only follow the cited medium. `all $\times$ @FoxNews excl.' is the cross similarity between the exclusive followers of @FoxNews in our dataset and all users (including the followers of @FoxNews) in our dataset.
                Top panel: Self-similarities of the different sub-populations   along with the cross-similarity  of exclusive followers of @FoxNews against the set of all users.
                Bottom panel: Same data recomputed after the suppression of the    \#covid topic from the topic vectors.}
        \end{figure}

   The top panel of Fig.~\ref{fig:similarity_nyt_fox} shows two remarkable  peaks  in  the self similarity curves, one by the end of March 2020 which corresponds to New York city's lockdown  and the other by the end of October 2020, the exception being  the self similarity of the curve of exclusive followers of @FoxNews, which has only one.   The bottom panel, shows the self-similarity recomputed suppressing the ``COVID'' topic from the topic vectors and the disappearing of the peak of March 2020 confirms that the synchronization of the discussion corresponds to this event. 
   
   Due to the large overlap of followers of different media, illustrated in Fig.~\ref{fig:venn}, it is not  surprising that the self-similarity curves of non-exclusive followers of different media show  a qualitatively similar behaviour. However, scrutinizing the exclusive followers of @nytimes and the exclusive followers of @FoxNews  we  observe they behave differently.

   When    the ``COVID'' component has been removed from the topic vectors of the users, the self similarity of the exclusive followers of @FoxNews is  higher than that of the rest of the users (including that of the exclusive followers of @nytimes), except for the large peak at the end of October that we will discuss later.  
    Remarkably, the top panel shows that  while the followers of @nytimes undergo the  synchronization period related to ``COVID'' topic, those of @FoxNews on the contrary, decrease their similarity, indicating that the ``COVID'' topic does not act as a synchronizing event for them.
   
   It should not be concluded that  exclusive @FoxNews followers at this time do not talk about the \#covid topic, but rather that the selection of topics they talk about becomes more inhomogeneous.
   In fact, we have found that the covid topic is not the most used one of  this subset of users (see Supplementary Material) .

 The very high peak of the end of October is present in  all the  curves in Fig.~\ref{fig:similarity_nyt_fox}, it corresponds to  the \#endsars topic, mentioned above, and it disappears when the corresponding topic is suppressed from the topic vectors.

       The cross-similarity curves between exclusive followers of  @FoxNews and a randomized sample of all users, is near zero  most of the time, and becomes negative, around the end of March, where all the other self similarities were increasing.  After July and before the  the \#endsars related peak mentioned above, the cross similarity approaches zero and so do all the self-similarities curves, with the exception of  the @FoxNews exclusive followers. Showing again that  those users talk, in general, about the same topics in the same terms, regardless the external events that may drive the attention of other users.

    \section{Discussion}
    \label{sec:discussion}
    
 We have  studied the dynamics of interactions  between the information agenda of a traditional medium, The New York Times, and the discussions that its followers  hold on Twitter. We also compare   with the discussion held by the followers of other  media  among the most followed  in U.S. involving TV news chains, newspapers, bi-weekly magazines, and press agencies.

Building   a semantic network of hashtags  with the only assumption  that two hashtags  used in the same tweet  refer to the same subject, we are able to automatically detect the topics discussed in Twitter by 
 community detection in this network.   For the NYT journal, the topics  are identified by the keywords chosen by the journalists to label their articles.

The entropy of hashtags and keywords usages captures  the structural differences among these two kind of media: the    curves of the entropy of the vocabulary used by the followers of all the media in Twitter show  very similar dynamics including  minor details, but all of them show a dynamical behaviour that is different  from that of the NYT journal. 

We observe that  the journal is much more concerned  with political news than \textit{its own followers}, as shown by the sudden decrease of keyword entropy located around key political dates,  for example, during the electoral period. 

Our results show that the entropy of the vocabulary of the set of  @FoxNews followers is significantly lower than for  any other media at any time.

Regarding  the agenda setting question, a relevant signal is found around the hashtag \#Blacklivesmatter, referring to the killing  of a black citizen during a  police intervention. We show that this discussion was originated on line and was treated by the journal short afterwards. 

The analysis of rank diversity of hashtags and keywords  uncovers a counter intuitive result: instead of finding the first ranks  completely dominated by the few forms of  COVID-19 hashtags in Twitter a high variability of the used hashtags dominates, and only the two first ranks have relatively low variability, which is nevertheless high enough so as to contain hundreds of hasthags. The situation is completely different for the journal which shows a slowly growing rank diversity of keywords, starting by very low values. This difference is expected as keywords, unlike hashtags, are curated and correspond to the sections of the journal that obey to a hierarchical order. Interestingly, the rank diversity in Twitter is also very different from that observed in Weibo (the Chinese version of Twitter)~\cite{cui2021attention}, which looks more like the rank diversity in the journal where keywords are curated.

     The  interaction between the journal and its followers has also been  explored by studying the patterns observed in the distribution of  time delays  of direct and indirect responses of the followers, to the articles and tweets posted by the journal. The main observation is the broad spectrum spanned by the time delays of the responses going from seconds up to a week, which may be surprising given the continuous flow of posts in Twitter. 
     
Similar heavy tail behaviour has been identified in studies of the distribution of delays in cascading processes~\cite{lu2014frequency,zhao2015seismic}, where the models  proposed to explain these patterns, mainly combine preferential attachment mechanisms with queuing processes~\cite{mathews2017nature,crane2008robust}.  However,  here we identify a similar distribution of response times in a different setup: instead of following a single cascading effect  triggered by an initial seed, which requires for the source tweet  to be detected by the users who will  potentially retweet (hence the preferential attachment mechanism proposed), we study the  behaviour of users who are in principle, automatically  exposed to each of the source tweets because they have decided to follow the journal's account. This questions the pertinence of  the preferential attachment hypothesis  to explain this observed pattern.

    On the contrary, the extremely similar behavior of all direct interactions suggests that this process may  strongly be influenced by a queuing process in the users' timeline, where older tweets might be pushed out quickly  by newer tweets. In this case, the first retweets of an article may trigger more retweets from the users that might have lost the original tweet from @nytimes in their timeline, in a manner of a self exciting process~\cite{hawkes1971spectra,rizoiu2017tutorial}.
  
  It is not  straightforward to foresee a single general hypothesis to explain the heavy tailed shape of the delay times distribution. A detailed analysis conditioned on the section of the NYT in which the articles were published, shows a dependence of the delay times on the sections, suggesting that some types of news have longer lifetimes than others. 
  
  On the other hand, our analysis of indirect reactions, where users post  tweets containing links to articles of the NYT, i.e., by clicking the `share via Twitter' button on the NYT website, shows reaction times that are as expected,  much slower. 
          
Finally,  the dynamical similarity  among groups of users allows to detect that, while most of the users synchronize their discussions around the date of lockdown, a singular behaviour is observed for exclusive followers  of @nytimes and of @FoxNews. The similarity of the former, although increasing in this period,  is  sensitively lower than the similarity of the global population, while for the latter, it shows in this period, the only long lasting  decrease of similarity (about a month). 

The relatively high and constant values of  similarity (except for the large peak related to \#endsars)  along with the low entropy of the vocabulary of the  exclusive followers of @FoxNews strongly suggest that this group constitutes an echo chamber. 
Moreover the cross-similarity among exclusive users of @nytimes and @FoxNews, is almost always negative (except for the singular \#endsars peak), which is  an objective measurement  of the strong separation of the subjects of interests of these two groups.

 \section{Conclusion}   
    
We present a dynamical study of the interactions between a traditional medium, the NYT journal and its followers in Twitter and we compare with the behaviour of Twitter users who follow other media of different kinds (written press, television, and press agencies).     It is important to stress that  we are not interested in the behaviour of a random sample of Twitter users but we are focusing instead on Twitter users that are interested in news, who could be thought to be a priori more susceptible to media influence. 
    
    Our results show     that as long as the users follow different media,  the similarity among  them is almost independent of the media sources they follow. On the contrary, the similarity becomes significantly different when observing  sub-populations of \textit{exclusive users}, those who follow one medium account exclusively. 
    We also show that this difference between sub-populations is dominant  around the first wave of COVID in the U.S. which in spite of being a public health topic that affects all populations, induces a differential behaviour on  sub-populations who exclusively follow different media.

    One important feature of our study is the fact that we avoid introducing selection bias by  choosing a priori some group of words. Here we keep the whole discussion as it is and we let the topics to emerge from  the communities detection process on the semantic networks. 
    
   Finally, we cannot stress enough the importance of choosing different independent quantities to analyse the data: it is the combination of the entropy of vocabulary with the similarity among the users which allows to  objectively show the singularity of the exclusive followers of @FoxNews with respect to the baseline population. In the same way, the comparison of the dynamics of entropy, topic evolution, and similarity, shows that although \#elections is a hot topic for the journal, the synchronization of its followers  around it, although measurable, is relatively lower compared with the \#Blacklivesmatter topic.  Moreover,    in spite of the general difficulty of detecting causality, the comparison of the dynamics of entropy and topic evolution, shows that the latter originated on Twitter before being treated by the NYT.

In summary, we present an automatic detection  method of discussion topics on social networks, which along with a set of independent measures on the obtained data,  bring  a lot of information with a minimum of assumptions (here the semantic link among  hashtags and among keywords), and should be the entrance gate to more detailed analysis that could focus on the treatment of specific topics or the detailed behaviour of specific groups.

       \section{Methods}

        In this section we present the data set used in this work, explaining the rationale leading to this particular choice, along with the procedure used for its  collection from different data sources. We define the semantic networks built with these data and we explain how we automatically detect the set of topics under discussion and the evolution of the attention each user pays to them.
        
        Moreover we also give the mathematical definition of  the observables used to characterize the dynamics of discussion in Twitter and that of the treatment of the news by the NYT over the studied period.

    \label{sec:model_methods}
    \subsection{Data collection}
    \subsubsection{Data from Twitter}
    
      \begin{table}[h]
            \begin{ruledtabular}
                \begin{tabular}{llr}
           
  @nytimes    &     Total collected users &    8'151'587 \\
         &  Total collected tweets & 502'647'015  \\
         &  Number of tweets with \# & 83'237'523  \\
        &   Number of distinct \# & 12'937'293    \\
         &  Number of users quoting/rt/reply & 226'630 \\
           
    \hline \noalign{\vskip 0.1cm}
   
   Other media  & Total collected users &   1'771'170 \\
               &  Total collected tweets & 96'551'331 \\
               
                \end{tabular}
            \end{ruledtabular}
            \caption{Data collected from Twitter. Top panel: random sample of the about 46M  followers of the NYT.Bottom panel followers of the other media described in the text.   }
        \label{Twitter_dat_nyt}
        \end{table}

We first recall briefly  the standard vocabulary used to name different elements of the  Twitter micro-blogging platform.   
       Users can engage on many different levels with each other. Each user has a \emph{Twitter handle}, which starts with `@'. They can write \emph{tweets}, short messages consisting of up to 280 characters, which may also contain images, videos or sound, and which are shown to their \emph{followers} -other users subscribing to the their accounts- on their \emph{timeline}, the list of latest posted tweets. However, even non-followers can see and interact with them (except for \emph{private tweets} which are not part of our dataset). Users can \emph{retweet} the tweet of another user, which means that they share this tweet with all  their followers. They can \emph{quote} a tweet, meaning that they republish the original tweet with a comment. Finally, they can \emph{reply} to a tweet, which starts a discussion connected to the original tweet. Tweets may contain 
       \emph{hashtags}, which are arbitrary strings of characters prefixed by the character `\#' ,  often used to tag the tweet.  Tweets can contain a \emph{URL}, which typically links to an external website. 

Due to the  very large number of followers (about 46 million) of the @nytimes, the official Twitter account of the NYT,  we have chosen for this study a random sample of them,  according to the following procedure:         
  \begin{itemize}
      \item We first obtained the list of the \emph{user ids} of all followers of @nytimes,  using the Twitter's official REST API \footnote{\emph{Representational state transfer application programming interface}, a common type of interface to web services}. This list was collected over a few days in the last week of June 2020. 
      \item We   randomized the obtained list.
      \item On July 1st 2020, we requested up to the last 3200 tweets (this number is a limitation of the Twitter API) of a sample of about 8M  of  these accounts. 
      \item Roughly every 2 months we requested, for all  users in our sample for which we already found tweets for the year 2020, the new tweets they published since our last query. 
  \end{itemize}

    Table~\ref{Twitter_dat_nyt} gives the main characteristics of the data used for this study.

   At the beginning of March 2021, we had collected  up to half a billion tweets published by more than 8M (8151587) followers.
   
    As it is well known that only a minority of Twitter users include their geolocalization in their profile, we have chosen not to control for this variable so as to avoid  artificially diminishing the number of collected users. However, since  the US is the largest market  both for Twitter and NYT, we expect that most followers are indeed located in US. As a consequence, although we cannot rule out that the dataset contains tweets of users living abroad,  we will naturally focus on  events that are relevant to the US in order to tag  the chronology of the study. The pertinence of this choice is supported by the fact that  topics which are popular in the US are dominating the discussion, and we  show that it is possible to identify the rare exceptions.

        This dataset, in the form of user and tweet ids, is available at \cite{rawDataNYTimes}.

        Although our  method enables us to collect a large  sampling of a specific subpopulation of Twitter, avoiding biases  that may be introduced by filtering, for example by hashtags, we discuss below  some limitations that might still remain in this data set, along with an estimation of  their potential influence in our study.
        \begin{itemize}
            \item     Due to the limit set by the API (it delivers only the last 3200 Tweets of the requested user), we risk to systematically miss tweets of very active accounts: those who would have tweeted more than 3200 tweets between January 1st and July 1st or those who would have exceeded that limit during the  $\sim 2$ month period of each collection step hold after July 1st 2020 until the end. Although most of such  accounts are  automated  (\emph{bots}) or institutional ones, like @nytimes itself,  one cannot rule out a priori, the existence of accounts of very active individuals. Notice that such users  need to write  at least  about 18 tweets per day, on average, in the first six months and many more in the  following collection periods (every two months), which is certainly possible but not typical of the standard user. 
        
     Nevertheless, in order to evaluate to what extent  our sample is likely to contain \textit{incomplete users} -accounts for which we could not get the full set of the content they published-  we set a conservative criterion to detect them. We count  the number of users for which we collected more than  3000 tweets. This strict bound leaves a generous room for deleted tweets, which although not downloaded, still count against the 3200 limit.
        Since we can collect at most 3200 tweets at each point in time, we can not exclude a priori, that a user wrote all these tweets and even more during one of our collections cycles (and  few or none in the other cycles).          However, we do not observe such inhomogeneous behavior, in spite of the fact that our sample contains  users who exceed  18000 tweets in all the period.         We are therefore confident that the strict  bound  set here overestimates the fraction of incomplete users considerably. According to this strict criterion,  we estimate that only less than $ 0.4\%$ of all accounts are incomplete  such that the induced error should be small, in particular considering that  our study makes a stronger usage of  the number of users rather than  the number of tweets. 
        
       \item The list of followers was fixed at the beginning of the study, such that we do not include users which started following @nytimes after July 1st 2020, slightly underestimating the influence of new and short lived accounts. 
       
       \item In the same way we can not exclude that some accounts we sampled stopped following @nytimes sometime during our period of study. 
       \item  Naturally, we do not consider in our sample  tweets from deleted, suspended and private accounts.
        \end{itemize}
        
        Following a similar technique, we also collected  a smaller sample of about a million users  who do not follow @nytimes but who follow instead, at least one of other seven   most followed US  news media accounts.  We do not include followers of  secondary accounts for example, those of ``breaking news'', like  @CNNbreaking. 
       Table~\ref{table_media} describes the different sources from where we have collected the sample of Twitter users interested in US news that we have studied in this work.
        
\begin{table}
\begin{ruledtabular}
\begin{tabular}{llr} 
 Name & Media type & followers \\ 
 \hline \noalign{\vskip 0.1cm}
 CNN & TV news  & 53'242'242 \\ 
 FoxNews & TV news & 20'121'721 \\
 Reuters & news agency & 23'238'148 \\ 
 Associated Press & news agency & 15'127'593 \\ 
 TIME & bi-weekly magazine & 18'065'949 \\
 Wall Street Journal & newspaper & 18'705'760 \\ 
 The Washington Post & newspaper & 17'791'609 \\ 
 The New York Times & newspaper & 46'808'154\\
\end{tabular}
\end{ruledtabular}
\caption{Number of followers of the Twitter accounts of the studied media.}\label{table_media}
\end{table}

        We collected a uniform sample of these users proportional to the number of followers each medium has, in the last weeks of March 2021. This means that the problem of missing tweets from very active accounts is worse for this data set. However, the fraction of incomplete accounts remains small $< 0.3\%$ (even smaller than for the @nytimes dataset, because we only had one cycle causing fewer false positives). Again, this dataset in the form of user and tweet ids is available at \cite{rawDataMedia}.
        
         Finally, we also collected all tweets of @nytimes account of the period, referenced by retweets, quotes or replies of their followers.
        
        In this study we only use metadata of the tweets: hashtags normalized to lower case (i.e., we treat \#covid-19 and \#COVID-19 as the same hashtag) and URLs. We do not extract further data from the remainder of the tweet, neither text nor images nor videos. Nevertheless, we will show that this minimalist information contained in the tweets already provides a rich image of the public discussion in the platform.  

\subsubsection{Data from the NYT}

Table~\ref{nyt} describes the main figures involved in the analysis of the publications of the NYT journal during the same period. 
        
    \begin{table}[h]
            \begin{ruledtabular}
                \begin{tabular}{lr}
           
    Number of articles  &  62'138  \\
    Number of tweets posted by @nytimes & 33'446  \\
    Links to  articles in @nytimes tweets &   20'496 \\
          Number of distinct keywords & 45'016 \\
                \end{tabular}
            \end{ruledtabular}
            \caption{Data concerning  the publications of the NYT journal in the considered period. }
        \label{nyt}
        \end{table}
        
   In addition  to the data from Twitter users, we collected the metadata of all articles published by the NYT either in print or online using their \emph{archive API}. This dataset includes in particular, a set of keywords for each article, which lists subjects, persons and locations referred to in the article. Moreover, it provides unique identifiers, which we used to connect URLs encountered in tweets, to a NYT article, an otherwise non trivial task, since an article can have multiple valid URLs. 
   
   The dataset that indicates which tweets link to which articles is also available at \cite{rawDataNYTimes}.
        
    \subsection{Observables}
       We detail in this section the  quantities  or \textit{observables}, that we  used in this study.
        
    \subsubsection{Entropy}
    \label{sec:def_entropy}
        The entropy of the hashtag distribution over a timeframe $t$ is defined as: 
        \begin{align}
            \label{eq:entropy}
            S_t = -\sum_i p_t(i) \ln p_t(i).
        \end{align}
        
        Where $p_t(i)$ is the probability distribution calculated as the ratio of the number of unique users that have used hashtag $i$ and the number of different pairs (hashtag, user)  within the time frame $t$.
       By considering unique users we   diminish the influence of very active accounts (e.g., spammers).  We calculate the entropy daily with a rolling time frame of seven days to remove the well known influence of the lower activity on weekends. 
    
    \subsubsection{Topic detection}
  In this study we are interested in comparing the dynamics of subjects published by a traditional medium, like the NYT, where  professionals  choose the information to be issued, with   
  the dynamics of discussion that its followers hold on the  Twitter platform. To do so one needs to identify the \emph{topics} that are discussed  in both media. In Twitter, we could use hashtags, which are used to tag the message, as a proxy for the subject that the tweet is about. However, multiple hashtags may address the same topic. A common strategy to follow the discussion about a topic is to pre-select the hashtags that are supposed to be related to the topic. Here we use a different approach where the topics emerge from a \emph{semantic network} of hashtags~\cite{cardoso2019topical,arg_elections}.        The vertices of this network are the  hashtags found in our dataset, and the weighted  edge between two nodes represents the number of different users that used those hashtags together in at least one of their tweets. In clear, if the same user publishes many tweets including the same pair of hashtags, it contributes to the weight only once. Finally, we set a  threshold for the link to be meaningful and we prune all edges whose weight is below 10. The rationale behind this construction is that two hashtags used in the same tweet refer to the same subject. 
        In this way, hashtags talking about the same topic should be strongly connected and synonymous hashtags, which only seldom appear in the same tweet, should be strongly connected to the same common nodes. 
        
        By performing community detection on the semantic network, we  detect the groups of hashtags that are more tightly  connected among them than with the rest  \cite{Fortunato2010community}. We identify each community with a topic of discussion in the platform. 
        
        This topic-community identification may suffer from some ambiguities  because some hashtags can belong to multiple topics. For example, if we use OSLOM2 \cite{Lancichinetti2011Finding} for community detection, which allows for community overlap, we find that  \#covid19 which influences most aspects of life,  is associated with more than 10 communities.
              Since overlapping communities are hard to interpret, we finally chose  a community detection algorithm, \emph{Infomap} \cite{Rosvall2008}, which provides a disjoint partition.  In this case \#covid19 will be assigned to one topic. 
        To illustrate the density of this network, a small fraction of it (the induced subgraphs of $\approx 1.5\%$ of the most co-used hashtags) is represented in Fig.~S5 of  the Supplementary Material.

        For keywords obtained from NYT articles, we do not need to perform such a topic analysis, since they are manually curated to already describe topics.
        
    \subsubsection{Rank diversity}
        The \emph{rank} $r$ of an entity (here either hashtags or keywords) is its position in the list of all entities occurring within a time period sorted by decreasing number of usages.
       Following \cite{cocho2015rank,Morales2018rank,cui2021attention}, we define the \emph{rank diversity} $d(r)$ over a time frame $\Delta$ with a time resolution $\delta$ as the number of of different entities occupying rank $r$ over the $k = \Delta / \delta$ time spans normalized by $k$. It therefore can assume values in $[1/k, 1]$, where $1/k$ signals that only a single entity was observed on the corresponding rank and $1$ that the entity changed for every period. Here, we study the $\Delta = 366$ days of 2020 and use a resolution of $\delta = 1$ day (starting at 0:00 UTC).
        
        This measures how consistent topics of interests are. Low values signal little fluctuations in the importance of the entities, while high values suggest high fluctuation. If $d(r)$ increases with the rank $r$, it signals that the really important topics are more consistent than minor topics. A decrease could happen if the entities are artificially curated, e.g., limited to a certain number.
        
    \subsubsection{User similarity}
    To study how similar users are in regard to the interest they pay to different topics,
        we applied the method used in  \cite{arg_elections}. 
        We describe the interests of each user $i$
        by means of a user description vector $\boldsymbol{d_i}$ of dimension $N_T$, 
        the number of topics (communities) found, which informs about the topic preferences of user $i$.

        This description vector is computed in the following way: 
        \begin{enumerate}
            \item  We build a user-topic matrix, $U$, where each element, $u_{ij}$, gives the absolute number of times that user $i$ has used a hashtag that belongs to the community identified as topic $j$. 
            \item  We compute the global topic vector $\boldsymbol{T}=\sum_i ^{N}{\boldsymbol{u_i}}$, where $\boldsymbol{u_i}$ is the $i$-th row vector in the user-topic matrix, and $N$ the size of the population. This vector gives the total number of times that each topic has been used by all the users in the dataset.
            \item  We define the vector $\boldsymbol{v_i}$ which gives the difference between the frequency of usage of the topic by user $i$ and its global frequency of usage in the population.
            
            \begin{equation}
               \boldsymbol{v_i}  = \frac{\boldsymbol{u_i}}{||\boldsymbol{u_i}||_1} - \frac{\boldsymbol{T}}{||\boldsymbol {T}||_1} \enspace.
               \label{frequency_vector}
            \end{equation}
            
            Here the norm $||.||_1$ must be understood as the sum over all the components in the space of dimension $N_T$. The vectors of Eq.~\ref{frequency_vector} thus inform about whether user $i$ has addressed each of the identified topics more or less than on average.
            
            \item As we are only interested in the orientation of the description vectors, they are normalized as:  
            \begin{equation}
                \boldsymbol{d_i} = \frac{\boldsymbol{v_i}}{||\boldsymbol{v_i}||_2} \enspace,
            \end{equation}
            where $||\boldsymbol{v_i}||_2$ is the standard euclidean norm in the topic hyperspace of dimension $N_T$.

        \end{enumerate}
        
        Then, in order to track the evolution of the users' interests we apply the aforementioned procedure to sliding time windows of 7 days, thus producing a series of matrices $U_t$, one for each day. We shall call $\boldsymbol{d_i}^t$  the description vector for user $i$ at discrete time $t$.
        
        We define the similarity between a pair of users $i$ and $j$ as the cosine similarity between the corresponding description vectors. As the latter are normalized, the similarity reduces to the inner product:
        \begin{equation}
            s(i,j) = \langle \boldsymbol{d_i} , \boldsymbol{d_j} \rangle \enspace.
        \end{equation}
        
        We also define the average description vector of a group of users $G$, of cardinality $|G|$: 
         
        \begin{equation}
           \boldsymbol{D_G} = \frac{\sum_{i \in G} \boldsymbol{d_i}}{|G|} \enspace.
        \end{equation}
        
        
        Now we can introduce two indices measuring \textit{collective similarities}:
        
        \begin{itemize}
            \item The \textit{cohesion} of a group of users, \textit{intra-group similarity} or \textit{self-similarity}, $s(G,G)$, defined as the average similarity between all its users, and computed in the following way:
        
            \begin{align}
                s(G,G) &= 
                   \frac{\sum_{i,j \in G}
                     {s(i,j)}}{|G|^2}
                    = \frac{\sum_{i \in G}{\langle \boldsymbol{d_i}, \boldsymbol{D_G}  \rangle}}{|G|}\\
                    &={\langle \boldsymbol{D_G}, \boldsymbol{D_G} \rangle}=||\boldsymbol{D_G}||^2 \enspace,
            \end{align}
            

            
            \item The \textit{cross-group similarity} is the average similarity between members of different groups $G_1$ and $G_2$, namely $s(G_1, G_2)$:
                
            \begin{align}
                s(G_1,G_2) &= 
                   \frac{\sum_{i\in G_1,j \in G_2}
                     {s(i,j)}}{|G_1|\cdot|G_2|} \\
                    &= \frac{\sum_{i \in G_1}{\langle \boldsymbol{d_i}, \boldsymbol{D_{G_2}} \rangle}}{|G_1|}={\langle \boldsymbol{D_{G_1}}, \boldsymbol{D_{G_2}} \rangle} \enspace.
            \end{align}
            
            
            
        \end{itemize}
        
    
     \section*{  \textbf{Data  Availability} }

     Dataset of user and tweet ids of followers of @nytimes is available at:
     \href{https://doi.org/10.5281/zenodo.4736651}{doi.org/10.5281/zenodo.4736651}.\\

Dataset of user and tweet ids of followers of other news  media  is available at: 
\href{https://doi.org/10.5281/zenodo.4736816}{doi.org/10.5281/zenodo.4736816}

 \section*{ \textbf{Author Contributions} }

 L.H., M.G.B., J.I.A.H proposed the research questions and the methodology, H.S. collected, curated, processed data, H.S. and M.G.B wrote the code for processing data, H.S., L.H., M.G.B., J.I.A.H, and D.K. performed result analysis,  L.H., M.G.B and H.S wrote the manuscript,  H.S., L.H., M.G.B., J.I.A.H, and D.K.  revised the manuscript. \\

 \section* {\textbf{Competing interests}} 

The authors declare no competing interests.\\

    \section*{Acknowledgments}
        The authors acknowledge the OpLaDyn grant obtained in the 4th round
        of the Trans-Atlantic Platform Digging into Data Challenge (2016-147 ANR OPLADYN TAP-DD2016).
        H.S. acknowledges grant Labex MME-DII (Grant No. ANR reference 11-LABEX-0023). J.I.A.H. and M.G.B. acknowledge the financial support of UBACyT-2018 20020170100421BA and the OpLaDyn grant HJ-253570 Annex IF-2017-14123506-APN-DNCEII\#MCT.

    \bibliography{lit}

\begin{thebibliography}{10}

\bibitem{10.2307/1019970}
William Hard.
\newblock Radio and public opinion.
\newblock {\em The Annals of the American Academy of Political and Social
  Science}, 177:105--113, 1935.

\bibitem{248510}
{La Réforme et le livre : l'Europe de l'imprimé (1517-v. 1570), dossier
  conçu et rassemblé par Jean-François Gilmont. Paris : Les Editions du
  Cerf, 1990. In-8°, 531 pages, illustrations. (Cerf-Histoire.)}.
\newblock 1991.

\bibitem{url_1899}
\url{https://www.nytimes.com/1899/05/07/archives/future-of-wireless-telegraphy.html}.

\bibitem{10.2307/2991719}
Susan~J. Douglas.
\newblock Public radio and television in america: A political history.
\newblock {\em The Public Opinion Quarterly}, 63(3):439--441, 1999.

\bibitem{gaumont2018reconstruction}
Noé Gaumont, Maziyar Panahi, and David Chavalarias.
\newblock {Reconstruction of the socio-semantic dynamics of political activist
  Twitter networks—Method and application to the 2017 French presidential
  election}.
\newblock {\em PLOS ONE}, 13(9):1--38, 09 2018.

\bibitem{boutet2013s}
Antoine Boutet, Hyoungshick Kim, and Eiko Yoneki.
\newblock {What’s in Twitter, I know what parties are popular and who you are
  supporting now!}
\newblock {\em Social network analysis and mining}, 3(4):1379--1391, 2013.

\bibitem{himelboim2013tweeting}
Itai Himelboim, Marc Smith, and Ben Shneiderman.
\newblock {Tweeting Apart: Applying Network Analysis to Detect Selective
  Exposure Clusters in Twitter}.
\newblock {\em Communication Methods and Measures}, 7(3-4):195--223, 2013.

\bibitem{barbera2015birds}
Pablo Barber{\'a}.
\newblock {Birds of the same feather tweet together: Bayesian ideal point
  estimation using Twitter data}.
\newblock {\em Political analysis}, 23(1):76--91, 2015.

\bibitem{ScienceOpenVid:2cb18dc9-e556-4c25-bcae-c0751724fde6}
Dimitar Nikolov, Diego~FM Oliveira, Alessandro Flammini, and Filippo Menczer.
\newblock Measuring online social bubbles.
\newblock {\em PeerJ computer science}, 1:e38, 2015.

\bibitem{cinelli2021echo}
Matteo Cinelli, Gianmarco De~Francisci Morales, Alessandro Galeazzi, Walter
  Quattrociocchi, and Michele Starnini.
\newblock The echo chamber effect on social media.
\newblock {\em Proceedings of the National Academy of Sciences},
  118(9):e2023301118, 2021.

\bibitem{choi2020rumor}
Daejin Choi, Selin Chun, Hyunchul Oh, Jinyoung Han, et~al.
\newblock Rumor propagation is amplified by echo chambers in social media.
\newblock {\em Scientific reports}, 10(1):1--10, 2020.

\bibitem{Lazer_fake}
David M.~J. Lazer, Matthew~A. Baum, Yochai Benkler, Adam~J. Berinsky, Kelly~M.
  Greenhill, Filippo Menczer, Miriam~J. Metzger, Brendan Nyhan, Gordon
  Pennycook, David Rothschild, Michael Schudson, Steven~A. Sloman, Cass~R.
  Sunstein, Emily~A. Thorson, Duncan~J. Watts, and Jonathan~L. Zittrain.
\newblock The science of fake news.
\newblock {\em Science}, 359(6380):1094--1096, 2018.

\bibitem{infodemics}
Riccardo Gallotti, Francesco Valle, Nicola Castaldo, Pierluigi Sacco, and
  Manlio De~Domenico.
\newblock {Assessing the risks of ‘infodemics’ in response to COVID-19
  epidemics}.
\newblock {\em Nature Human Behaviour}, 4(12):1285--1293, 2020.

\bibitem{menczer2021covid}
Kai-Cheng Yang, Francesco Pierri, Pik-Mai Hui, David Axelrod, Christopher
  Torres-Lugo, John Bryden, and Filippo Menczer.
\newblock {The COVID-19 Infodemic: Twitter versus Facebook}.
\newblock {\em Big Data \& Society}, 8(1):20539517211013861, 2021.

\bibitem{shahi2021exploratory}
Gautam~Kishore Shahi, Anne Dirkson, and Tim~A. Majchrzak.
\newblock {An exploratory study of COVID-19 misinformation on Twitter}.
\newblock {\em Online Social Networks and Media}, 22:100104, 2021.

\bibitem{cohen2015press}
Bernard~Cecil Cohen.
\newblock {\em Press and foreign policy}.
\newblock princeton university press, 2015.

\bibitem{10.1086/267990}
Maxwell~E McCombs and Donald~L Shaw.
\newblock The agenda-setting function of mass media.
\newblock {\em Public opinion quarterly}, 36(2):176--187, 1972.

\bibitem{SACCO2021114215}
Pier~Luigi Sacco, Riccardo Gallotti, Federico Pilati, Nicola Castaldo, and
  Manlio De~Domenico.
\newblock {Emergence of knowledge communities and information centralization
  during the COVID-19 pandemic}.
\newblock {\em Social Science \& Medicine}, 285:114215, 2021.

\bibitem{cinelli2020covid}
Matteo Cinelli, Walter Quattrociocchi, Alessandro Galeazzi, Carlo~Michele
  Valensise, Emanuele Brugnoli, Ana~Lucia Schmidt, Paola Zola, Fabiana Zollo,
  and Antonio Scala.
\newblock {The COVID-19 social media infodemic}.
\newblock {\em Scientific Reports}, 10:16598, 2020.

\bibitem{collection_covid}
Emilio Ferrara, Stefano Cresci, and Luca Luceri.
\newblock Misinformation, manipulation, and abuse on social media in the era of
  covid-19.
\newblock {\em Journal of Computational Social Science}, 3(2):271--277, 2020.

\bibitem{vargo2015event}
Chris~J Vargo, Ekaterina Basilaia, and Donald~Lewis Shaw.
\newblock {Event versus issue: Twitter reflections of major news, a case
  study}.
\newblock {\em Studies in Media and Communications}, 9:215--239, 2015.

\bibitem{morris2018twitter}
David~S. Morris.
\newblock Twitter versus the traditional media: A survey experiment comparing
  public perceptions of campaign messages in the 2016 u.s. presidential
  election.
\newblock {\em Social Science Computer Review}, 36(4):456--468, 2018.

\bibitem{bridgman2021infodemic}
Aengus Bridgman, Eric Merkley, Oleg Zhilin, Peter~John Loewen, Taylor Owen, and
  Derek Ruths.
\newblock {Infodemic Pathways: Evaluating the Role That Traditional and Social
  Media Play in Cross-National Information Transfer}.
\newblock {\em Frontiers in Political Science}, 3:20, 2021.

\bibitem{su2019agenda}
Yan Su and Porismita Borah.
\newblock Who is the agenda setter? examining the intermedia agenda-setting
  effect between twitter and newspapers.
\newblock {\em Journal of Information Technology \& Politics}, 16(3):236--249,
  2019.

\bibitem{ceron2014twitter}
Andrea Ceron.
\newblock Twitter and the traditional media: Who is the real agenda setter?
\newblock In {\em APSA 2014 Annual Meeting Paper}, 2014.

\bibitem{arg_elections}
Tom{\'a}s~Mussi Reyero, Mariano~G Beir{\'o}, J~Ignacio Alvarez-Hamelin, Laura
  Hern{\'a}ndez, and Dimitris Kotzinos.
\newblock Evolution of the political opinion landscape during electoral
  periods.
\newblock {\em EPJ Data Science}, 10(1):31, 2021.

\bibitem{cardoso2019topical}
Felipe~Maciel Cardoso, Sandro Meloni, Andr{\'e} Santanch{\`e}, and Yamir
  Moreno.
\newblock Topical alignment in online social systems.
\newblock {\em Frontiers in Physics}, 7:58, 2019.

\bibitem{cocho2015rank}
Germinal Cocho, Jorge Flores, Carlos Gershenson, Carlos Pineda, and Sergio
  Sánchez.
\newblock {Rank Diversity of Languages: Generic Behavior in Computational
  Linguistics}.
\newblock {\em PLoS One}, 10(4):e0121898, 04 2015.

\bibitem{Morales2018rank}
José~A. Morales, Ewan Colman, Sergio Sánchez, Fernanda Sánchez-Puig, Carlos
  Pineda, Gerardo Iñiguez, Germinal Cocho, Jorge Flores, and Carlos
  Gershenson.
\newblock {Rank Dynamics of Word Usage at Multiple Scales}.
\newblock {\em Frontiers in Physics}, 6:45, 2018.

\bibitem{hawkes1971spectra}
ALAN~G. HAWKES.
\newblock {Spectra of some self-exciting and mutually exciting point
  processes}.
\newblock {\em Biometrika}, 58(1):83--90, 04 1971.

\bibitem{rizoiu2017tutorial}
Marian-Andrei Rizoiu, Young Lee, Swapnil Mishra, and Lexing Xie.
\newblock {}a tutorial on hawkes processes for events in social media.

\bibitem{young2015everything}
P.~Young.
\newblock {\em Everything You Wanted to Know About Data Analysis and Fitting
  but Were Afraid to Ask}.
\newblock SpringerBriefs in Physics. Springer International Publishing, 2015.

\bibitem{cui2021attention}
{Cui, Hao} and {Kert\'esz, J\'anos}.
\newblock {Attention dynamics on the Chinese social media Sina Weibo during the
  COVID-19 pandemic}.
\newblock {\em EPJ Data Sci.}, 10(1):8, 2021.

\bibitem{lu2014frequency}
Yao Lu, Peng Zhang, Yanan Cao, Yue Hu, and Li~Guo.
\newblock On the frequency distribution of retweets.
\newblock {\em Procedia Computer Science}, 31:747--753, 2014.
\newblock 2nd International Conference on Information Technology and
  Quantitative Management, ITQM 2014.

\bibitem{zhao2015seismic}
Qingyuan Zhao, Murat~A. Erdogdu, Hera~Y. He, Anand Rajaraman, and Jure
  Leskovec.
\newblock Seismic: A self-exciting point process model for predicting tweet
  popularity.
\newblock In {\em Proceedings of the 21th ACM SIGKDD International Conference
  on Knowledge Discovery and Data Mining}, KDD '15, page 1513–1522, New York,
  NY, USA, 2015. Association for Computing Machinery.

\bibitem{mathews2017nature}
Peter Mathews, Lewis Mitchell, Giang Nguyen, and Nigel Bean.
\newblock The nature and origin of heavy tails in retweet activity.
\newblock In {\em Proceedings of the 26th International Conference on World
  Wide Web Companion}, WWW '17 Companion, page 1493–1498, Republic and Canton
  of Geneva, CHE, 2017. International World Wide Web Conferences Steering
  Committee.

\bibitem{crane2008robust}
Riley Crane and Didier Sornette.
\newblock Robust dynamic classes revealed by measuring the response function of
  a social system.
\newblock {\em Proceedings of the National Academy of Sciences},
  105(41):15649--15653, 2008.

\bibitem{rawDataNYTimes}
Hendrik Schawe.
\newblock {Dataset of user and tweet ids of followers of @nytimes }, May 2021.
\newblock https://doi.org/10.5281/zenodo.4736651.

\bibitem{rawDataMedia}
Hendrik Schawe.
\newblock {Dataset of user and tweet ids of followers of news outlet media
  accounts }, May 2021.
\newblock https://doi.org/10.5281/zenodo.4736816.

\bibitem{Fortunato2010community}
Santo Fortunato.
\newblock Community detection in graphs.
\newblock {\em Physics Reports}, 486(3):75--174, 2010.

\bibitem{Lancichinetti2011Finding}
Andrea Lancichinetti, Filippo Radicchi, José~J. Ramasco, and Santo Fortunato.
\newblock Finding statistically significant communities in networks.
\newblock {\em PLOS ONE}, 6(4):1--18, 04 2011.

\bibitem{Rosvall2008}
Martin Rosvall and Carl~T. Bergstrom.
\newblock Maps of random walks on complex networks reveal community structure.
\newblock {\em Proceedings of the National Academy of Sciences},
  105(4):1118--1123, 2008.

\end{thebibliography}

\end{document}